\documentclass[american]{article}
\usepackage[T1]{fontenc}
\usepackage[latin9]{inputenc}
\usepackage{geometry}
\usepackage{color}
\usepackage{babel}
\usepackage{calc}
\usepackage{textcomp}
\usepackage{mathrsfs}
\usepackage{amsmath}
\usepackage{amssymb}
\usepackage{graphicx}
\usepackage[unicode=true]
 {hyperref}

\makeatletter

\newcommand{\lyxaddress}[1]{
\par {\raggedright #1
\vspace{1.4em}
\noindent\par}
}

\makeatother

\begin{document}

\title{Ultra-short Pulse Propagation in Nonlinear Optics}

\author{Andrés Macho Ortiz and Roberto Llorente Sáez}
\maketitle

\lyxaddress{\textcolor{black}{\small{}Nanophotonics Technology Center, Universitat
Politècnica de València, Camino de Vera s/n, 46022 Valencia, Spain.
(\href{mailto:amachor@ntc.upv.es}{amachor@ntc.upv.es})}}
\begin{abstract}
In this work, we perform a detailed review about the theoretical modeling
of the optical propagation of ultra-short pulses ($\sim$ femtoseconds)
in different scenarios of nonlinear optics. In particular, we focus
our efforts on optical fibers and uniform media, with special attention
on the white light continuum generation by using gases.
\end{abstract}
\vspace{1cm}

\tableofcontents{}

\pagebreak{}

\section{Introduction\label{sec:1}}

The propagation of electromagnetic waves can be modeled by using Maxwell's
equations, the essence of classical electromagnetism. They are able
to describe the structure of the electromagnetic field and the fundamental
interactions between the electric field strength ($\boldsymbol{\mathcal{E}}$)
and the magnetic induction ($\boldsymbol{\mathcal{B}}$). In their
fundamental form, the \emph{macroscopic} Maxwell equations read as
follows (we use the notation $\partial_{t}=\partial/\partial t$):
\begin{align}
\nabla\times\boldsymbol{\mathcal{E}}\left(\boldsymbol{r},t\right) & =-\partial_{t}\boldsymbol{\mathcal{B}}\left(\boldsymbol{r},t\right);\label{eq:1.1}\\
c_{0}^{2}\nabla\times\boldsymbol{\mathcal{B}}\left(\boldsymbol{r},t\right) & =\frac{1}{\varepsilon_{0}}\boldsymbol{\mathcal{J}}\left(\boldsymbol{r},t\right)+\partial_{t}\boldsymbol{\mathcal{E}}\left(\boldsymbol{r},t\right);\label{eq:1.2}\\
\varepsilon_{0}\nabla\cdot\boldsymbol{\mathcal{E}}\left(\boldsymbol{r},t\right) & =\rho\left(\boldsymbol{r},t\right);\label{eq:1.3}\\
\nabla\cdot\boldsymbol{\mathcal{B}}\left(\boldsymbol{r},t\right) & =0.\label{eq:1.4}
\end{align}

\noindent where $\rho$ and $\boldsymbol{\mathcal{J}}$ are the total
charge and current densities, respectively\footnote{\noindent All the fields appearing in Eqs.\,(\ref{eq:1.1})-(\ref{eq:1.4})
are volume-averaged quantities, expressed in S.I. units.}\textsuperscript{,}\footnote{\noindent $c_{0}=1/\sqrt{\varepsilon_{0}\mu_{0}}$ is the speed of
light in vacuum. $\varepsilon_{0}$ and $\mu_{0}$ are respectively
the electric permittivity and the magnetic permeability in vacuum.}. Both terms include the free and bound charges and currents, i.e.,
$\rho=\rho_{\textrm{f}}+\rho_{\textrm{b}}$ and $\boldsymbol{\mathcal{J}}=\boldsymbol{\mathcal{J}}_{\textrm{f}}+\boldsymbol{\mathcal{J}}_{\textrm{b}}$
\cite{key-1}.

As is well known, the fields $\boldsymbol{\mathcal{E}}$ and $\boldsymbol{\mathcal{B}}$
are able to modify the charge distribution of a medium and induce
currents in it. Such alterations act as new sources that generate
additional electromagnetic fields. Thus, it is necessary to know the
electromagnetic response of the medium to model theoretically the
different electromagnetic wave propagation phenomena. These field-matter
interactions can be adequately described by defining two new auxiliary
macroscopic fields, namely $\boldsymbol{\mathcal{D}}$ (electric displacement)
and $\boldsymbol{\mathcal{H}}$ (magnetic field strength), which account
for the macroscopic response of the medium charges and currents to
the applied fields. The connection between the fundamental and auxiliary
fields is given through the so-called constitutive relations: 
\begin{equation}
\begin{alignedat}{1}\boldsymbol{\mathcal{D}}\left(\boldsymbol{r},t\right):=\varepsilon_{0}\boldsymbol{\mathcal{E}}\left(\boldsymbol{r},t\right)+\boldsymbol{\mathcal{P}}\left(\boldsymbol{r},t\right);\ \ \  & \boldsymbol{\mathcal{H}}\left(\boldsymbol{r},t\right):=\frac{1}{\mu_{0}}\boldsymbol{\mathcal{B}}\left(\boldsymbol{r},t\right)-\boldsymbol{\mathcal{M}}\left(\boldsymbol{r},t\right)\end{alignedat}
,\label{eq:1.5}
\end{equation}

\noindent where $\boldsymbol{\mathcal{P}}$ and $\boldsymbol{\mathcal{M}}$
are the polarization and magnetization fields, and basically express
the density of electric and magnetic dipole moments, respectively.

In a dielectric medium (e.g. an optical fiber), the free charges and
currents are found to be null, $\rho=\rho_{\textrm{b}}=-\nabla\cdot\boldsymbol{\mathcal{P}}$,
$\boldsymbol{\mathcal{J}}=\boldsymbol{\mathcal{J}}_{\textrm{b}}=\partial_{t}\boldsymbol{\mathcal{P}}$,
$\boldsymbol{\mathcal{M}}=\mathbf{0}$ and Eqs.\,(\ref{eq:1.1})-(\ref{eq:1.4})
are reduced to\textcolor{red}{{} }\cite{key-1}:\footnote{Eqs.\,(\ref{eq:1.6})-(\ref{eq:1.9}) are also valid to describe
the structure of the electromagnetic field in vacuum by taking $\boldsymbol{\mathcal{P}}=\boldsymbol{0}$.} 
\begin{align}
\nabla\times\boldsymbol{\mathcal{E}}\left(\boldsymbol{r},t\right) & =-\partial_{t}\boldsymbol{\mathcal{B}}\left(\boldsymbol{r},t\right);\label{eq:1.6}\\
\nabla\times\boldsymbol{\mathcal{H}}\left(\boldsymbol{r},t\right) & =\partial_{t}\boldsymbol{\mathcal{D}}\left(\boldsymbol{r},t\right);\label{eq:1.7}\\
\nabla\cdot\boldsymbol{\mathcal{D}}\left(\boldsymbol{r},t\right) & =0;\label{eq:1.8}\\
\nabla\cdot\boldsymbol{\mathcal{B}}\left(\boldsymbol{r},t\right) & =0.\label{eq:1.9}
\end{align}
A rigorous electromagnetic analysis of the linear and nonlinear pulse
propagation through an optical fiber should be performed by solving
Eqs.\,(\ref{eq:1.6})-(\ref{eq:1.9}) in the temporal and longitudinal
orders of reference of $\boldsymbol{\mathcal{E}}$ and $\boldsymbol{\mathcal{B}}$,
i.e., $\delta t\sim2\pi/\omega_{0}$ and $\delta z\sim\lambda$, where
$\omega_{0}$ and $\lambda$ are the values of the angular frequency
and the wavelength of the laser in the core of the fiber. 

However, the numerical calculation of these equations requires a high
computational time and memory when several meters ($z>10^{6}\lambda$)
or kilometers ($z>10^{9}\lambda$) are involved. In such a case, it
is more suitable to consider the temporal ($T_{\textrm{P}}$) and
longitudinal width ($\Lambda_{\textrm{P}}$) of the optical pulses
as the temporal and longitudinal orders of reference ($\delta t\sim T_{\textrm{P}}$
and $\delta z\sim\Lambda_{\textrm{P}}$). In other words, the complexity
of the numerical calculations can be alleviated if we are able to
simulate only the slowly-varying temporal and longitudinal evolution
of $\boldsymbol{\mathcal{E}}$ and $\boldsymbol{\mathcal{B}}$, the
so-called \emph{complex envelope} ($\mathcal{A}$, see Fig.\,1). 
\noindent \begin{center}
\includegraphics[width=10cm,height=4cm,keepaspectratio]{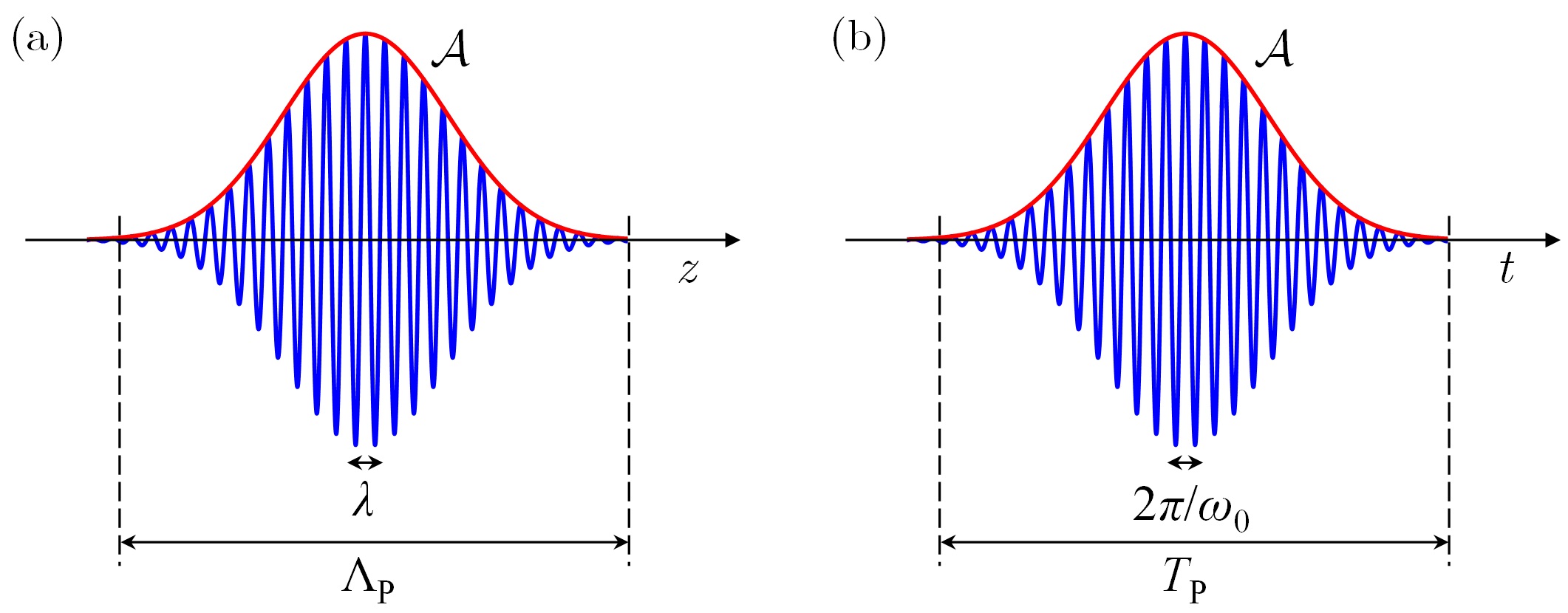}
\par\end{center}

\noindent {\footnotesize{}Fig. 1. Optical pulse: (a) longitudinal
and (b) temporal changes. (Blue line: rapidly-varying fluctuations.
Red line: slowly-varying fluctuations, complex envelope.)\\}{\footnotesize \par}

The partial differential equation modeling the linear and nonlinear
propagation of $\mathcal{A}$ in guided and unguided optical media
will be referred to as the pulse propagation equation (PPE) in this
work. In the photonics literature, this equation is usually termed
as the generalized paraxial wave equation or the pulse propagation
equation \cite{key-2,key-3,key-4}. Specifically, in guided media,
it can also be referred to as the nonlinear Schrödinger equation (NLSE)
due to the intimate relation between the PPE and the NLSE when the
space and time are normalized \cite{key-5}. In the following, we
first review the mathematical derivation of the PPE from Maxwell's
equations when propagating femtosecond optical pulses through a single-core
single-mode optical fiber, later, we discuss the propagation in uniform
($\equiv$ unguided) nonlinear dielectric media, and finally, we focus
our attention in gases for white light continuum generation applications.

\section{Pulse propagation in optical fibers\label{sec:2}}

In general, the derivation of the PPE is simpler than initially foreseen.
Figure 2 shows a flowchart of the steps required during our mathematical
discussion. Let us remember that our goal is to derive a partial differential
equation of the complex envelope (the slowly-varying longitudinal
and temporal evolution of the electromagnetic field) starting from
Maxwell's equations.
\noindent \begin{center}
\includegraphics[width=14.5cm,height=13cm,keepaspectratio]{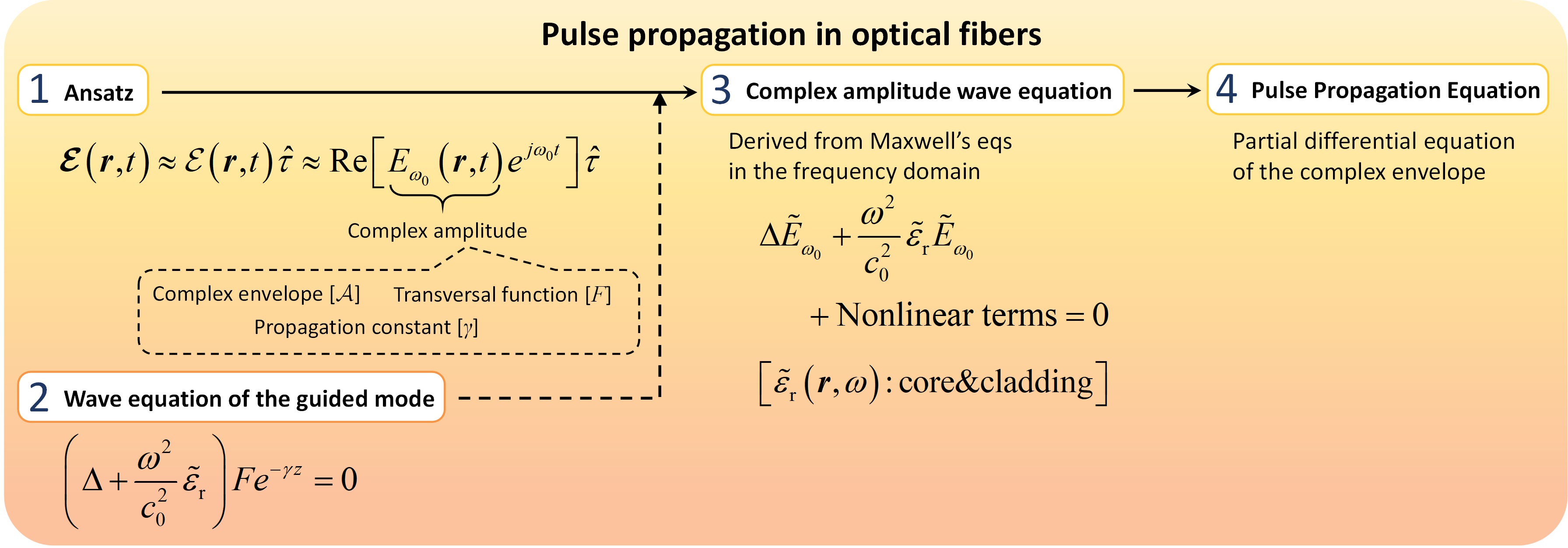}{\footnotesize{}}\\
{\footnotesize{}Fig. 2. Flowchart of transformations to derive the
PPE from Maxwell's equations in optical fibers.}
\par\end{center}{\footnotesize \par}

\pagebreak{}

\subsubsection*{First step: Ansatz of the electric field strength of the fiber}

In the first step, we should propose the ansatz of the global electric
field strength ($\boldsymbol{\mathcal{E}}$) of the fiber\textcolor{black}{}\footnote{Since the electric field strength and the magnetic induction can be
expressed in identical form, we will use the former field for our
theoretical discussions with straightforward extrapolation of the
results to the magnetic induction by using the intrinsic impedance
of each dielectric region. }. In this work, we will consider an ultra-short optical pulse (femtosecond
regime) generated from a laser with optical carrier of angular frequency
$\omega_{0}$ and launched into a single-core single-mode fiber. For
the sake of simplicity, we will assume the weakly-guiding approximation
(also referred to as the paraxial approximation in the photonics literature)
and omit the polarization and birefringent effects of the optical
medium. Consequently, the longitudinal component of $\boldsymbol{\mathcal{E}}$
can be neglected and we can approximate $\boldsymbol{\mathcal{E}}\simeq\mathcal{E}\hat{\tau}$,
where $\hat{\tau}$ is a transverse unitary vector ($\hat{\tau}\perp\hat{z}$).
Without loss of generality, $\hat{\tau}$ can be related to the $x$
or $y$ axis.

At this point, we should decouple the rapidly and the slowly-varying
temporal and longitudinal changes of $\boldsymbol{\mathcal{E}}$.
The rapidly-varying \emph{temporal} changes are decoupled by using
the \emph{slowly-varying complex amplitude approximation} (SVAA):
\begin{equation}
\mathcal{E}\left(\boldsymbol{\boldsymbol{r}},t\right)\simeq\textrm{Re}\left\{ E_{\omega_{0}}\left(\boldsymbol{\boldsymbol{r}},t\right)\exp\left(j\omega_{0}t\right)\right\} =\frac{1}{2}\left[E_{\omega_{0}}\left(\boldsymbol{\boldsymbol{r}},t\right)\exp\left(j\omega_{0}t\right)+E_{-\omega_{0}}\left(\boldsymbol{\boldsymbol{r}},t\right)\exp\left(-j\omega_{0}t\right)\right],\label{eq:2.1}
\end{equation}
where $E_{\omega_{0}}$ is the complex amplitude of the electric field
strength of the fundamental mode $\textrm{LP}_{01}$ (polarized along
the $\tau$ axis) satisfying that $\left|\delta_{t}E_{\omega_{0}}\right|\ll\left|E_{\omega_{0}}\right|$
in $\delta t\sim2\pi/\omega_{0}$, where $\delta_{t}E_{\omega_{0}}:=E_{\omega_{0}}\left(\boldsymbol{\boldsymbol{r}},t+\delta t\right)-E_{\omega_{0}}\left(\boldsymbol{\boldsymbol{r}},t\right).$
It should be noted that the SVAA allows us to decouple the rapid temporal
oscillation of the optical carrier from the slow temporal evolution
of the complex amplitudes of the optical pulses {[}Fig.\,1(b), red
line{]}. Therefore, the herein proposed model is valid \emph{if and
only if} the Maxwell equations are approximately satisfied when using
Eq.\,(\ref{eq:2.1}). However, this assumption is not fulfilled if
the pulse is too narrow, namely around the order of the period of
the optical carrier or shorter. In such a case, the decomposition
performed in Eq.\,(\ref{eq:2.1}) is no longer useful and the concept
of the complex amplitude itself becomes unclear. Specifically, the
validity of Eq.\,(\ref{eq:2.1}) can be easily tested by verifying
that the pulse bandwidth is much lower than $\omega_{0}/2\pi$. 

Now, we should also decouple the rapidly- and slowly-varying \emph{longitudinal}
variations of the electric field {[}Fig.\,1(a), red line{]} by introducing
the complex envelope in Eq.\,(\ref{eq:2.1}). Let us describe this
point in detail:
\begin{enumerate}
\item First, for simplicity, we can start by assuming a laser radiating
a continuous-wave (CW) signal. In this case, we can write the complex
amplitude $E_{\omega_{0}}$ as follows: 
\begin{align}
E_{\omega_{0}}\left(\boldsymbol{\boldsymbol{r}},t\right) & =\frac{1}{2\pi}\int_{-\infty}^{\infty}\widetilde{\mathtt{A}}\thinspace F\left(x,y,\omega\right)\exp\left(-\gamma\left(\omega\right)z\right)\exp\left(j\left(\omega-\omega_{0}\right)t\right)\textrm{d}\omega,\label{eq:2.2}
\end{align}
where $\widetilde{\mathtt{A}}$ is a complex constant related to the
optical power of the CW signal ($P\propto\left|\widetilde{\mathtt{A}}\right|^{2}$),
$F$ is the transverse Bessel eigenfunction of the $\textrm{LP}_{01}$
mode \cite{key-6} and $\gamma=\alpha/2+j\beta$ is the propagation
constant involving the power attenuation constant $\alpha$ and the
phase constant $\beta$. Note that in Eq.\,(\ref{eq:2.2}) we have
only considered the forward propagation by using the sign $-$ by
convention. We can omit the backward propagation because we are are
not considering fiber perturbations along the $z$ axis and the optical
fiber is assumed with a low nonlinear nature. In such a case, we can
neglect the reflected power \cite{key-7}. Thus, the Fourier transform\footnote{The direct and inverse Fourier transforms employed in this work are
defined as: 
\[
\widetilde{X}\left(\omega\right)=\mathcal{F}\left[X\left(t\right)\right]:=\int_{-\infty}^{\infty}X\left(t\right)\exp\left(-j\omega t\right)\textrm{d}t;\ \ \ \ X\left(t\right)=\mathcal{F}^{-1}\left[\widetilde{X}\left(\omega\right)\right]:=\frac{1}{2\pi}\int_{-\infty}^{\infty}\widetilde{X}\left(\omega\right)\exp\left(j\omega t\right)\textrm{d}\omega.
\]
} of Eq.\,(\ref{eq:2.1}) is found to be: 
\begin{align}
\widetilde{\boldsymbol{\mathcal{E}}}\left(\boldsymbol{\boldsymbol{r}},\omega\right) & =\mathcal{F}\left[\boldsymbol{\mathcal{E}}\left(\boldsymbol{\boldsymbol{r}},t\right)\right]\simeq\textrm{Re}\left\{ \widetilde{E}_{\omega_{0}}\left(\boldsymbol{\boldsymbol{r}},\omega-\omega_{0}\right)\right\} \hat{\tau},\label{eq:2.3}
\end{align}
where:
\begin{equation}
\widetilde{E}_{\omega_{0}}\left(\boldsymbol{\boldsymbol{r}},\omega-\omega_{0}\right)=\mathcal{F}\left[E_{\omega_{0}}\left(\boldsymbol{\boldsymbol{r}},t\right)\exp\left(j\omega_{0}t\right)\right]=\widetilde{\mathtt{A}}\thinspace F\left(x,y,\omega\right)\exp\left(-\gamma\left(\omega\right)z\right).\label{eq:2.4}
\end{equation}
Equation\,(\ref{eq:2.4}) is the well-known solution of an $\textrm{LP}$
mode in the classical modal analysis of a single-core fiber \cite{key-6}.
\item Now, assume a pulsed-wave laser. In this case, the constant $\widetilde{\mathtt{A}}$
is found to be frequency-dependent accounting for the amplitude of
each spectral component of the pulse propagated by the $\textrm{LP}_{01}$
mode. That is, Eq.\,(\ref{eq:2.4}) should be rewritten as: 
\begin{equation}
\widetilde{E}_{\omega_{0}}\left(\boldsymbol{\boldsymbol{r}},\omega-\omega_{0}\right)=\widetilde{\mathtt{A}}\left(\omega-\omega_{0}\right)F\left(x,y,\omega\right)\exp\left(-\gamma\left(\omega\right)z\right),\label{eq:2.5}
\end{equation}
with the complex function $\widetilde{\mathtt{A}}$ modeling the Fourier
transform of the optical pulse in \emph{baseband}. Concretely, in
Eq.\,(\ref{eq:2.5}), $\widetilde{\mathtt{A}}$ is shifted to the
angular frequency of the optical carrier $\omega_{0}$ because $\widetilde{E}_{\omega_{0}}$
(the slowly-varying temporal evolution of the electric field) is shifted
to $\omega_{0}$.
\item In the pulsed-wave scenario of the previous point, $\widetilde{\mathtt{A}}$
is the complex envelope of the optical pulse in the frequency domain.
Unfortunately, $\widetilde{\mathtt{A}}$ is only able to describe
the slowly-varying temporal evolution of the electric field because
is the shape of the optical pulse at $z=0$. To include the slowly-varying
\emph{spatial} evolution of the electric field at any $z$ point we
should multiply this function by an exponential of the form $\exp\left(-j\beta\left(\omega\right)z\right)$
but subtracting the rapidly-varying longitudinal variations of the
optical carrier $\exp\left(-j\beta\left(\omega_{0}\right)z\right)$.
That is, we should perform the transformation: 
\begin{equation}
\widetilde{A}\left(z,\omega-\omega_{0}\right):=\widetilde{\mathtt{A}}\left(\omega-\omega_{0}\right)\exp\left[-j\left(\beta\left(\omega\right)-\beta\left(\omega_{0}\right)\right)z\right].\label{eq:2.6}
\end{equation}
Note that the subtraction $\beta\left(\omega\right)-\beta\left(\omega_{0}\right)$
in the spatial exponential is analog to the subtraction $\omega-\omega_{0}$
in the temporal exponential of Eq.\,(\ref{eq:2.2}). Remarkably,
the new complex envelope $\widetilde{A}$ describes, in the frequency
domain, the slowly-varying \emph{longitudinal} and \emph{temporal}
evolution of the global electric field of the fiber ($\boldsymbol{\mathcal{E}}$).
\end{enumerate}
All in all, the ansatz of Maxwell's equations (first step of Fig.\,2)
is described by the following expressions in the time domain {[}$\beta\left(\omega_{0}\right)\equiv\beta_{0}${]}:
\begin{align}
\mathcal{E}\left(\boldsymbol{\boldsymbol{r}},t\right) & \simeq\textrm{Re}\left\{ E_{\omega_{0}}\left(\boldsymbol{\boldsymbol{r}},t\right)\exp\left(j\omega_{0}t\right)\right\} ;\label{eq:2.7}\\
E_{\omega_{0}}\left(\boldsymbol{\boldsymbol{r}},t\right) & =\frac{1}{2\pi}\int_{-\infty}^{\infty}\widetilde{A}\left(z,\omega-\omega_{0}\right)F\left(x,y,\omega\right)\exp\left(-j\beta_{0}z-\frac{1}{2}\alpha\left(\omega\right)z\right)\exp\left(j\left(\omega-\omega_{0}\right)t\right)\textrm{d}\omega,\label{eq:2.8}
\end{align}
and in the frequency domain:
\begin{align}
\widetilde{\boldsymbol{\mathcal{E}}}\left(\boldsymbol{\boldsymbol{r}},\omega\right) & \simeq\textrm{Re}\left\{ \widetilde{E}_{\omega_{0}}\left(\boldsymbol{\boldsymbol{r}},\omega-\omega_{0}\right)\right\} \hat{\tau};\label{eq:2.9}\\
\widetilde{E}_{\omega_{0}}\left(\boldsymbol{\boldsymbol{r}},\omega-\omega_{0}\right) & =\widetilde{A}\left(z,\omega-\omega_{0}\right)F\left(x,y,\omega\right)\exp\left(-j\beta_{0}z-\frac{1}{2}\alpha\left(\omega\right)z\right),\label{eq:2.10}
\end{align}
with the complex envelope in the time domain ($A$) given by the inverse
Fourier transform of $\widetilde{A}$: 
\begin{equation}
A\left(z,t\right)=\mathcal{F}^{-1}\left[\widetilde{A}\left(z,\Omega\right)\right]=\frac{1}{2\pi}\int_{-\infty}^{\infty}\widetilde{A}\left(z,\Omega\right)\exp\left(j\Omega t\right)\textrm{d}\Omega,\label{eq:2.11}
\end{equation}
and satisfying the slowly-varying envelope approximation (SVEA) in
the space and time domain. Mathematically, the SVEA can be expressed
as $\left|\delta_{z}A\right|\ll\left|A\right|$ in $\delta z\sim\lambda=2\pi/\beta_{0}$
and $\left|\delta_{t}A\right|\ll\left|A\right|$ in $\delta t\sim2\pi/\omega_{0}$,
where $\delta_{z}A:=A(z+\delta z,t)-A(z,t)$ and $\delta_{t}A:=A(z,t+\delta t)-A(z,t)$.
From the above statement, we found the order criteria of the SVEA:
\begin{align}
\left|\partial_{z}^{2}A\right|\ll\beta_{0}\left|\partial_{z}A\right|\ll\beta_{0}^{2}\left|A\right|; & \ \ \ \left|\partial_{t}^{2}A\right|\ll\omega_{0}\left|\partial_{t}A\right|\ll\omega_{0}^{2}\left|A\right|.\label{eq:2.12}
\end{align}
It should be noted that the temporal order criterion of the SVEA is
fulfilled thanks to the SVAA, performed in Eq.\,(\ref{eq:2.1}).
Moreover, it is noticeable that we are using a different typography
to describe the complex envelope ($A$) from the typography depicted
in Figs.\,1 and 2 ($\mathcal{A}$). Later, a new complex envelope
transformation $\mathcal{A}:=A\exp\left(-\alpha z/2\right)$ will
be performed to include the attenuation constant $\alpha$ in the
left-hand side (LHS) of the PPE and isolating the nonlinear terms
(sources) in the right-hand side (RHS).

\subsubsection*{Second step: Wave equation of the guided mode}

From the classical modal analysis of a single-core fiber \cite{key-6},
it is worthy to note that the guided mode $F\exp(-\gamma\left(\omega\right)z)$
should satisfy the Helmholtz equation in each spectral component $\omega$:
\begin{equation}
\left(\triangle+\frac{\omega^{2}}{c_{0}^{2}}\widetilde{\varepsilon}_{\textrm{r}}\left(\boldsymbol{\boldsymbol{r}},\omega\right)\right)F\left(x,y,\omega\right)\exp\left(-\gamma\left(\omega\right)z\right)=0,\label{eq:2.13}
\end{equation}
where $\widetilde{\varepsilon}_{\textrm{r}}$ is the Fourier transform
of the relative electric permittivity of the fiber, including the
core and cladding regions {[}see below Eq.\,(\ref{eq:2.24}) for
more details{]}. In particular, the above equation is the step 2 of
Fig.\,2.

\subsubsection*{Third step: Nonlinear complex amplitude wave equation}

In the third step, we should derive the nonlinear wave equation of
the optical medium as a function of the complex amplitude in the frequency
domain ($\widetilde{E}_{\omega_{0}}$). We start by combining the
Faraday\textquoteright s and Ampère\textquoteright s laws {[}Eqs.\,(\ref{eq:1.6})
and (\ref{eq:1.7}){]}. Using the constitutive relations given by
Eq.\,(\ref{eq:1.5}), the nonlinear wave equation of the electric
field strength is found to be: 
\begin{equation}
\triangle\boldsymbol{\mathcal{E}}\left(\boldsymbol{\boldsymbol{r}},t\right)-\frac{1}{c_{0}^{2}}\partial_{t}^{2}\boldsymbol{\mathcal{E}}\left(\boldsymbol{\boldsymbol{r}},t\right)=\mu_{0}\partial_{t}^{2}\boldsymbol{\mathcal{P}}^{\left(1\right)}\left(\boldsymbol{\boldsymbol{r}},t\right)+\mu_{0}\partial_{t}^{2}\boldsymbol{\mathcal{P}}^{\left(3\right)}\left(\boldsymbol{\boldsymbol{r}},t\right),\label{eq:2.14}
\end{equation}
where $\boldsymbol{\mathcal{P}}^{\left(1\right)}$ and $\boldsymbol{\mathcal{P}}^{\left(3\right)}$
are the linear and nonlinear polarization of the fiber, respectively.
Note that we have neglected the term $\nabla(\nabla\cdot(\boldsymbol{\mathcal{P}}^{\left(1\right)}+\boldsymbol{\mathcal{P}}^{\left(3\right)}))$
in Eq.\,(\ref{eq:2.14}) considering slowly-varying refractive index
profiles\footnote{That is, $\delta_{r}n\ll n$ in $\delta r\sim\lambda$, where $\delta_{r}n:=\left|n\left(r+\delta r\right)-n\left(r\right)\right|$.}
and the low birefringent and low nonlinear nature of silica fibers.
Unfortunately, this approximation has not been rigorously verified
in the literature \cite{key-2,key-3,key-4,key-5}. For the above reason,
we include in an appendix of \cite{key-7} a numerical verification
of this point.

Now, applying the Fourier transform to Eq.\,(\ref{eq:2.14}) we obtain:
\begin{equation}
\left(\triangle+\frac{\omega^{2}}{c_{0}^{2}}\right)\widetilde{\boldsymbol{\mathcal{E}}}\left(\boldsymbol{\boldsymbol{r}},\omega\right)=-\omega^{2}\mu_{0}\boldsymbol{\widetilde{\mathcal{P}}}^{\left(1\right)}\left(\boldsymbol{\boldsymbol{r}},\omega\right)-\omega^{2}\mu_{0}\boldsymbol{\widetilde{\mathcal{P}}}^{\left(3\right)}\left(\boldsymbol{\boldsymbol{r}},\omega\right).\label{eq:2.15}
\end{equation}
Remember that the field $\widetilde{\boldsymbol{\mathcal{E}}}$ can
be expressed in terms of $\widetilde{E}_{\omega_{0}}$ by using Eq.\,(\ref{eq:2.9}).
In a similar way, $\boldsymbol{\widetilde{\mathcal{P}}}^{\left(1\right)}$
and $\boldsymbol{\widetilde{\mathcal{P}}}^{\left(3\right)}$ can be
written in terms of their complex amplitude in the frequency domain
as: 
\begin{align}
\boldsymbol{\widetilde{\mathcal{P}}}^{\left(k\right)}\left(\boldsymbol{\boldsymbol{r}},\omega\right) & =\mathcal{F}\left[\boldsymbol{\mathcal{P}}^{\left(k\right)}\left(\boldsymbol{\boldsymbol{r}},t\right)\right]\simeq\mathcal{F}\left[\textrm{Re}\left\{ P_{\omega_{0}}^{\left(k\right)}\left(\boldsymbol{\boldsymbol{r}},t\right)\exp\left(j\omega_{0}t\right)\right\} \hat{\tau}\right]=\textrm{Re}\left\{ \widetilde{P}_{\omega_{0}}^{\left(k\right)}\left(\boldsymbol{\boldsymbol{r}},\omega-\omega_{0}\right)\right\} \hat{\tau}\nonumber \\
 & =\frac{1}{2}\left[\widetilde{P}_{\omega_{0}}^{\left(k\right)}\left(\boldsymbol{\boldsymbol{r}},\omega-\omega_{0}\right)+\widetilde{P}_{-\omega_{0}}^{\left(k\right)}\left(\boldsymbol{\boldsymbol{r}},\omega+\omega_{0}\right)\right]\hat{\tau};\ \ \ k\in\left\{ 1,3\right\} .\label{eq:2.16}
\end{align}
where the nonlinear polarization term in $3\omega_{0}$ was omitted
taking into account that the phase-matching condition in this nonlinear
term is not satisfied in silica fibers \cite{key-8}. Thus, Eq.\,(\ref{eq:2.15})
becomes: 
\begin{equation}
\left(\triangle+\frac{\omega^{2}}{c_{0}^{2}}\right)\widetilde{E}_{\omega_{0}}\left(\boldsymbol{\boldsymbol{r}},\omega-\omega_{0}\right)=-\omega^{2}\mu_{0}\widetilde{P}_{\omega_{0}}^{\left(1\right)}\left(\boldsymbol{\boldsymbol{r}},\omega-\omega_{0}\right)-\omega^{2}\mu_{0}\widetilde{P}_{\omega_{0}}^{\left(3\right)}\left(\boldsymbol{\boldsymbol{r}},\omega-\omega_{0}\right).\label{eq:2.17}
\end{equation}

Now, let us discuss the constitutive relation between $\widetilde{P}_{\omega_{0}}^{\left(1\right)}$
and $\widetilde{E}_{\omega_{0}}$. Modeling the linear response of
the fiber ($\boldsymbol{\mathcal{P}}^{\left(1\right)}$) to an incident
electric field ($\boldsymbol{\mathcal{E}}$) as a system: isotropic,
spatially heterogeneous nondispersive and temporally invariant and
dispersive, the linear polarization in the time domain $\boldsymbol{\mathcal{P}}^{\left(1\right)}$
can be expressed as:
\begin{equation}
\boldsymbol{\mathcal{P}}^{\left(1\right)}\left(\boldsymbol{\boldsymbol{r}},t\right)=\varepsilon_{0}\int_{-\infty}^{\infty}\chi^{\left(1\right)}\left(\boldsymbol{r},t-\zeta\right)\boldsymbol{\mathcal{E}}\left(\boldsymbol{\boldsymbol{r}},\zeta\right)\textrm{d}\zeta,\label{eq:2.18}
\end{equation}
with $\varepsilon_{0}$ the electric permittivity in vacuum and $\chi^{\left(1\right)}$
the first-order electric susceptibility. Hence, the linear polarization
can be regarded as a linear and time-invariant system, with the constitutive
relation of the linear polarization expressed in the frequency domain
as:
\begin{equation}
\boldsymbol{\widetilde{\mathcal{P}}}^{\left(1\right)}\left(\boldsymbol{\boldsymbol{r}},\omega\right)=\varepsilon_{0}\widetilde{\chi}^{\left(1\right)}\left(\boldsymbol{r},\omega\right)\widetilde{\boldsymbol{\mathcal{E}}}\left(\boldsymbol{\boldsymbol{r}},\omega\right),\label{eq:2.19}
\end{equation}
where:
\begin{equation}
\widetilde{\chi}^{\left(1\right)}\left(\boldsymbol{r},\omega\right):=\mathcal{F}\left[\chi^{\left(1\right)}\left(\boldsymbol{r},t\right)\right]=\int_{-\infty}^{\infty}\chi^{\left(1\right)}\left(\boldsymbol{r},t\right)\exp\left(-j\omega t\right)\textrm{d}t.\label{eq:2.20}
\end{equation}
It should be taken into account that the real part of $\widetilde{\chi}^{\left(1\right)}$
gives information about the material dispersion and the imaginary
part accounts for the optical absorption induced by the resonant frequencies
of fused silica, located in the ultraviolet (68.4 nm and 116.2 nm)
and infrared (9896.2 nm) bands \cite{key-2}. In the third transmission
window, the optical absorption is mainly induced by the Rayleigh scattering
(modeled by $\alpha$) and the imaginary part of $\widetilde{\chi}^{\left(1\right)}$
can be neglected. In any case, the complex amplitude of the linear
polarization in the frequency domain is found to be:
\begin{equation}
\widetilde{P}_{\omega_{0}}^{\left(1\right)}\left(\boldsymbol{\boldsymbol{r}},\omega-\omega_{0}\right)=\varepsilon_{0}\widetilde{\chi}^{\left(1\right)}\left(\boldsymbol{\boldsymbol{r}},\omega\right)\widetilde{E}_{\omega_{0}}\left(\boldsymbol{\boldsymbol{r}},\omega-\omega_{0}\right).\label{eq:2.21}
\end{equation}

Now, we still need to investigate the nonlinear polarization term
$\widetilde{P}_{\omega_{0}}^{\left(3\right)}$ of Eq.\,(\ref{eq:2.17}).
Nonetheless, the constitutive relation of the nonlinear polarization
with the electric field strength involves a convolution operation
in the frequency domain, increasing the complexity of the mathematical
discussion to derive the PPE. Hence, for simplicity, we will maintain
the term $\widetilde{P}_{\omega_{0}}^{\left(3\right)}$ in Eq.\,(\ref{eq:2.17}),
which becomes:
\begin{equation}
\left(\triangle+\frac{\omega^{2}}{c_{0}^{2}}\right)\widetilde{E}_{\omega_{0}}\left(\boldsymbol{\boldsymbol{r}},\omega-\omega_{0}\right)+\frac{\omega^{2}}{c_{0}^{2}}\widetilde{\chi}^{\left(1\right)}\left(\boldsymbol{\boldsymbol{r}},\omega\right)\widetilde{E}_{\omega_{0}}\left(\boldsymbol{\boldsymbol{r}},\omega-\omega_{0}\right)=-\omega^{2}\mu_{0}\widetilde{P}_{\omega_{0}}^{\left(3\right)}\left(\boldsymbol{\boldsymbol{r}},\omega-\omega_{0}\right).\label{eq:2.22}
\end{equation}
Furthermore, defining $\widetilde{\varepsilon}_{\textrm{r}}\left(\boldsymbol{\boldsymbol{r}},\omega\right):=1+\widetilde{\chi}^{\left(1\right)}\left(\boldsymbol{\boldsymbol{r}},\omega\right)$
and omitting the imaginary part of $\widetilde{\chi}^{\left(1\right)}$
as indicated before, the nonlinear complex amplitude wave equation
is finally derived: 
\begin{equation}
\left(\triangle+\frac{\omega^{2}}{c_{0}^{2}}\widetilde{\varepsilon}_{\textrm{r}}\left(\boldsymbol{\boldsymbol{r}},\omega\right)\right)\widetilde{E}_{\omega_{0}}\left(\boldsymbol{\boldsymbol{r}},\omega-\omega_{0}\right)=-\omega^{2}\mu_{0}\widetilde{P}_{\omega_{0}}^{\left(3\right)}\left(\boldsymbol{\boldsymbol{r}},\omega-\omega_{0}\right).\label{eq:2.23}
\end{equation}

\subsubsection*{Fourth step: Complex envelope propagation equation}

In the fourth step, we should finally derive the PPE by using the
results of step 1 {[}Eq.\,(\ref{eq:2.10}){]} and step 2 {[}Eq.\,(\ref{eq:2.13}){]}
in step 3 {[}Eq.\,(\ref{eq:2.23}){]}. However, the fourth step in
not straightforward. We should previously indicate some relevant points
about $\widetilde{\varepsilon}_{\textrm{r}}\left(\boldsymbol{\boldsymbol{r}},\omega\right)$
and $\beta\left(\omega\right)$.

In particular, $\widetilde{\varepsilon}_{\textrm{r}}\left(\boldsymbol{\boldsymbol{r}},\omega\right)$
gives information about the relative electric permittivity distribution
in the core and cladding regions. It is related with the refractive
index of the core ($n_{\textrm{co}}$) and cladding ($n_{\textrm{cl}}$)
as indicated below:
\begin{align}
\widetilde{\varepsilon}_{\textrm{r}}\left(\boldsymbol{\boldsymbol{r}},\omega\right) & =\widetilde{\varepsilon}_{\textrm{r},\textrm{cl}}\left(\boldsymbol{\boldsymbol{r}},\omega\right)+\Delta\widetilde{\varepsilon}_{\textrm{r},\textrm{co}}\left(\boldsymbol{\boldsymbol{r}},\omega\right)\nonumber \\
 & =\begin{cases}
\boldsymbol{\boldsymbol{r}}\equiv\textrm{core} & \widetilde{\varepsilon}_{\textrm{r},\textrm{co}}\left(\boldsymbol{\boldsymbol{r}},\omega\right)=\widetilde{\varepsilon}_{\textrm{r},\textrm{cl}}\left(\boldsymbol{\boldsymbol{r}},\omega\right)+\Delta\widetilde{\varepsilon}_{\textrm{r},\textrm{co}}\left(\boldsymbol{\boldsymbol{r}},\omega\right)\equiv n_{\textrm{co}}^{2}\left(\boldsymbol{\boldsymbol{r}},\omega\right)\\
\boldsymbol{\boldsymbol{r}}\equiv\textrm{cladding} & \,\widetilde{\varepsilon}_{\textrm{r},\textrm{cl}}\left(\boldsymbol{\boldsymbol{r}},\omega\right)\equiv n_{\textrm{cl}}^{2}\left(\boldsymbol{\boldsymbol{r}},\omega\right)
\end{cases},\label{eq:2.24}
\end{align}
where $\widetilde{\varepsilon}_{\textrm{r},\textrm{co}}$ and $\widetilde{\varepsilon}_{\textrm{r},\textrm{cl}}$
are respectively the relative electric permittivity in the core and
in the cladding, and $\Delta\widetilde{\varepsilon}_{\textrm{r},\textrm{co}}$
is the difference between $\widetilde{\varepsilon}_{\textrm{r},\textrm{co}}$
and $\widetilde{\varepsilon}_{\textrm{r},\textrm{cl}}$. Along this
line note that we have assumed these functions spatial dependent to
describe step and gradual-index fibers\footnote{The cladding can also be designed with a gradual-index profile (e.g.
hollow fibers)}. In addition, remember that $\widetilde{\varepsilon}_{\textrm{r}}$
is assumed to be a real function taking into account that we have
neglected the imaginary part of $\widetilde{\chi}^{\left(1\right)}$
when operating in the third transmission window.

On the other hand, it is worthy to note that Eq.\,(\ref{eq:2.13})
allows us to obtain a fundamental relation to describe the chromatic
dispersion. It is straightforward to derive this relation performing
the next two steps. First, the phase constant $\beta$ should be written
as: 
\begin{equation}
\beta\left(\omega\right)=\sum_{k=0}^{\infty}\frac{1}{k!}\left(\omega-\omega_{0}\right)^{k}\beta_{k},\label{eq:2.25}
\end{equation}
with $\beta_{k}:=\textrm{d}^{k}\beta(\omega=\omega_{0})/\textrm{d}\omega^{k}$.
\textcolor{black}{Second, we should approximate:}
\begin{align}
\beta^{2}\left(\omega\right)-\beta_{0}^{2} & =\left(\beta\left(\omega\right)+\beta_{0}\right)\left(\beta\left(\omega\right)-\beta_{0}\right)\nonumber \\
 & \simeq2\beta_{0}\left(\beta\left(\omega\right)-\beta_{0}\right)=2\beta_{0}\sum_{k=1}^{\infty}\frac{1}{k!}\left(\omega-\omega_{0}\right)^{k}\beta_{k}\equiv-j2\beta_{0}\mathfrak{D}\left(\omega\right),\label{eq:2.26}
\end{align}
where $\mathfrak{D}$ is the complex function defined as\textcolor{black}{:
}
\begin{equation}
\mathfrak{D}\left(\omega\right):=\sum_{k=1}^{\infty}\frac{j}{k!}\left(\omega-\omega_{0}\right)^{k}\beta_{k}.\label{eq:2.27}
\end{equation}
Thus, using Eq.\,(\ref{eq:2.26}) \textcolor{black}{we }directly
obtain from Eq.\,(\ref{eq:2.13}) {[}step 2{]} the expression that
allows us to describe the chromatic dispersion during the propagation
of the pulses: 
\begin{equation}
\left(\triangle_{\textrm{T}}+\frac{\omega^{2}}{c_{0}^{2}}\widetilde{\varepsilon}_{\textrm{r}}\left(\boldsymbol{\boldsymbol{r}},\omega\right)\right)F\left(x,y,\omega\right)=\left[\beta_{0}^{2}-j2\beta_{0}\mathfrak{D}\left(\omega\right)-j\alpha\left(\omega\right)\beta\left(\omega\right)\right]F\left(x,y,\omega\right),\label{eq:2.28}
\end{equation}
with $\triangle_{\textrm{T}}=\partial_{x}^{2}+\partial_{y}^{2}$ the
transverse Laplacian operator. 

Now, at this point we can start with the derivation of the PPE by
inserting our ansatz {[}Eq.\,(\ref{eq:2.10}){]} in the nonlinear
complex amplitude wave equation {[}Eq.\,(\ref{eq:2.23}){]} and using
Eq.\,(\ref{eq:2.28}) during the algebraic work. This is the solid
line connecting the steps 1 and 2 with step 3 which gives rise to
step 4 in Fig.\,2. \textcolor{black}{As a result, }after some algebraic
work and retaining the second-order longitudinal derivatives of the
complex envelopes considering that $\partial_{z}^{2}\widetilde{A}\neq0$
in $\delta z\sim\lambda$ for femtosecond optical pulses (Appendix
B4 of \cite{key-7}), we obtain (the independent variables are omitted
for simplicity):
\begin{align}
F\exp\left(-j\beta_{0}z\right)\left[\partial_{z}^{2}-\left(j2\beta_{0}+\alpha\right)\left(\partial_{z}+\mathfrak{D}\right)\right] & \widetilde{A}=-\omega^{2}\mu_{0}\widetilde{P}_{\omega_{0}}^{\left(3\right)}\exp\left(\frac{1}{2}\alpha z\right).\label{eq:2.29}
\end{align}
As can be seen, the LHS of Eq.\,(\ref{eq:2.29}) describes the linear
propagation and the RHS accounts for the nonlinear effects (sources).
In this vein, the derivation of the PPE requires to investigate the
constitutive relation between the complex amplitude of the nonlinear
polarization and the electric field strength in the frequency domain. 

To this end, we should start from the constitutive relation in the
time domain. Therefore, at this point let us remember the expressions
of the nonlinear polarization in the time and frequency domain as
a function of the corresponding complex amplitudes: 
\begin{align}
\boldsymbol{\mathcal{P}}^{\left(3\right)}\left(\boldsymbol{\boldsymbol{r}},t\right) & =\mathcal{P}^{\left(3\right)}\left(\boldsymbol{\boldsymbol{r}},t\right)\hat{\tau}\simeq\textrm{Re}\left\{ P_{\omega_{0}}^{\left(3\right)}\left(\boldsymbol{\boldsymbol{r}},t\right)\exp\left(j\omega_{0}t\right)\right\} \hat{\tau};\label{eq:2.30}\\
\boldsymbol{\widetilde{\mathcal{P}}}^{\left(3\right)}\left(\boldsymbol{\boldsymbol{r}},\omega\right) & =\mathcal{F}\left[\boldsymbol{\mathcal{P}}^{\left(3\right)}\left(\boldsymbol{\boldsymbol{r}},t\right)\right]\simeq\textrm{Re}\left\{ \widetilde{P}_{\omega_{0}}^{\left(3\right)}\left(\boldsymbol{\boldsymbol{r}},\omega-\omega_{0}\right)\right\} \hat{\tau},\label{eq:2.31}
\end{align}
with: 
\begin{equation}
\widetilde{P}_{\omega_{0}}^{\left(3\right)}\left(\boldsymbol{\boldsymbol{r}},\omega-\omega_{0}\right)=\mathcal{F}\left[P_{\omega_{0}}^{\left(3\right)}\left(\boldsymbol{\boldsymbol{r}},t\right)\exp\left(j\omega_{0}t\right)\right],\label{eq:2.32}
\end{equation}
and omitting the nonlinear polarization in $3\omega_{0}$, as indicated
before. Hence, the nonlinear terms of Eq.\,(\ref{eq:2.29}) can be
found by analyzing the constitutive relation $\boldsymbol{\mathcal{P}}^{\left(3\right)}$-\emph{$\boldsymbol{\mathcal{E}}$}
in the time domain and performing the Fourier transform of the complex
amplitudes.

In ultra-short optical pulses, the constitutive relation between the
nonlinear polarization and the electric field strength should include
the delay response of the electronic and nuclei structure of silica
atoms when an electric field stimulates the optical medium. The most
general expression to describe the third-order nonlinear response
in silica media is to consider an output of a system: nonlinear, anisotropic,
spatially homogeneous\footnote{Although the nonlinear electric susceptibility can be found slightly
different in each dielectric region of an optical fiber, we assume
spatial homogeneity in $\chi_{ijkl}^{\left(3\right)}$ due to the
low nonlinear nature of such optical media. In particular, we found
that $\chi_{ijkl}^{\left(3\right)}\sim10^{\text{\textminus}22}\ \textrm{m}^{2}/\textrm{V}^{2}$
in the core and cladding regions.} nondispersive, and temporally varying and dispersive. In such circumstances,
the constitutive relation can be written as {[}Einstein summation
convention, $\left(i,j,k,l\right)\in\left\{ x,y,z\right\} ^{4}${]}:
\begin{equation}
\mathcal{P}_{i}^{\left(3\right)}\left(\boldsymbol{\boldsymbol{r}},t\right)=\varepsilon_{0}\iiint_{\infty}\chi_{ijkl}^{\left(3\right)}\left(t,\zeta_{1},\zeta_{2},\zeta_{3}\right)\mathcal{E}_{j}\left(\boldsymbol{\boldsymbol{r}},\zeta_{1}\right)\mathcal{E}_{k}\left(\boldsymbol{\boldsymbol{r}},\zeta_{2}\right)\mathcal{E}_{l}\left(\boldsymbol{\boldsymbol{r}},\zeta_{3}\right)\textrm{d}\zeta_{1}\textrm{d}\zeta_{2}\textrm{d}\zeta_{3}.\label{eq:2.33}
\end{equation}
The time dispersive nature allows us to describe the exact time delay
induced by the electronic and nuclei response of silica atoms to the
incident electric field. For optical frequencies well below the electronic
transitions, the electronic contribution to the nonlinear polarization
can be considered instantaneous. However, since nucleons (protons
and neutrons) are considerably heavier than electrons, the nuclei
motions have resonant frequencies much lower than the electronic transitions
and, consequently, they should be retained in the constitutive relation.
Specifically, \emph{Raman scattering} is a well-known effect arising
from the nuclear contribution to the nonlinear polarization. Therefore,
considering ultra-short optical pulses wider than 1 fs, the electronic
response can be assumed instantaneous and Eq.\,(\ref{eq:2.33}) can
be reduced to \cite{key-3}:
\begin{align}
\mathcal{P}_{i}^{\left(3\right)}\left(\boldsymbol{\boldsymbol{r}},t\right) & \simeq\mathcal{P}_{i}^{\left(3\textrm{I}\right)}\left(\boldsymbol{\boldsymbol{r}},t\right)+\mathcal{P}_{i}^{\left(3\textrm{R}\right)}\left(\boldsymbol{\boldsymbol{r}},t\right)\nonumber \\
 & =\varepsilon_{0}\chi_{ijkl}^{\left(3\textrm{I}\right)}\left(t\right)\mathcal{E}_{j}\left(\boldsymbol{\boldsymbol{r}},t\right)\mathcal{E}_{k}\left(\boldsymbol{\boldsymbol{r}},t\right)\mathcal{E}_{l}\left(\boldsymbol{\boldsymbol{r}},t\right)+\varepsilon_{0}\mathcal{E}_{j}\left(\boldsymbol{\boldsymbol{r}},t\right)\int_{-\infty}^{\infty}\chi_{ijkl}^{\left(3\textrm{R}\right)}\left(t-\zeta\right)\mathcal{E}_{k}\left(\boldsymbol{\boldsymbol{r}},\zeta\right)\mathcal{E}_{l}\left(\boldsymbol{\boldsymbol{r}},\zeta\right)\textrm{d}\zeta,\label{eq:2.34}
\end{align}
where the first term of the RHS describes the instantaneous response
($3\textrm{I}$) accounting for the electronic resonances, and the
second term describes the nuclei motions inducing the intrapulse stimulated
Raman scattering effect ($3\textrm{R}$). The third-order electric
susceptibility tensors of the above equation can be expressed in silica
fibers as follows \cite{key-9}:
\begin{align}
\chi_{ijkl}^{\left(3\textrm{I}\right)}\left(t\right) & =\chi_{\textrm{NL}}\left(\frac{1-f_{\textrm{R}}}{3}\right)\left(\delta_{ij}\delta_{kl}+\delta_{ik}\delta_{jl}+\delta_{il}\delta_{jk}\right)\delta\left(t\right);\label{eq:2.35}\\
\chi_{ijkl}^{\left(3\textrm{R}\right)}\left(t\right) & =\chi_{\textrm{NL}}f_{\textrm{R}}\left[h\left(t\right)\delta_{ij}\delta_{kl}+\frac{1}{2}u\left(t\right)\left(\delta_{ik}\delta_{jl}+\delta_{il}\delta_{jk}\right)\right],\label{eq:2.36}
\end{align}
where $\delta_{ij}$ is the Kronecker delta function; $\chi_{\textrm{NL}}=2.6\text{·}10^{\text{\textminus}22}\ \textrm{m}^{2}/\textrm{V}^{2}$\textcolor{black}{{}
at the wavelength of 1550 nm}; $f_{\textrm{R}}=0.245$ represents
the fractional contribution of the delayed Raman response to the nonlinear
polarization; and $h$ and $u$ functions describe the isotropic and
anisotropic Raman response, respectively \cite{key-9}:
\begin{align}
h\left(t\right) & =f_{1}\tau_{1}\left(\tau_{1}^{-2}+\tau_{2}^{-2}\right)\exp\left(-t/\tau_{2}\right)\sin\left(t/\tau_{1}\right);\label{eq:2.37}\\
u\left(t\right) & =f_{2}\left(\frac{2\tau_{3}-t}{\tau_{3}^{2}}\right)\exp\left(-t/\tau_{3}\right)+\frac{f_{3}}{f_{1}}h\left(t\right),\label{eq:2.38}
\end{align}
with $f_{1}$, $f_{2}$, $f_{3}$, $\tau_{1}$, $\tau_{2}$ and $\tau_{3}$
constants of the nonlinear medium satisfying the following relations:
\begin{equation}
\begin{array}{ccc}
\chi_{\textrm{NL}}\left(1-f_{\textrm{R}}\right)=\frac{4}{3}\varepsilon_{0}c_{0}n_{\textrm{NL}}\overline{n}^{2}; & \sum_{i=1}^{3}f_{i}=1; & \int_{-\infty}^{\infty}\left(\chi_{ijkl}^{\left(3\textrm{I}\right)}\left(t\right)+\chi_{ijkl}^{\left(3\textrm{R}\right)}\left(t\right)\right)\textrm{d}t\end{array}=\chi_{\textrm{NL}},\label{eq:2.39}
\end{equation}
with $\overline{n}$ the average value of the material refractive
index of the fiber and $n_{\textrm{NL}}$ the nonlinear refractive
index modeling the changes induced in $\overline{n}$ by the nonlinear
polarization. In our case, we assume $\overline{n}\simeq1.45$ and
$n_{\textrm{NL}}=2.6\cdot10^{-20}\textrm{\ m}^{2}/\textrm{W}$. From
Eq.\,(\ref{eq:2.39}) and Ref.\,\cite{key-9} we found: $f_{1}=0.75$,
$f_{2}=0.21$, $f_{3}=0.04$, $\tau_{1}=12.2$ fs, $\tau_{2}=32$
fs, $\tau_{3}=96$ fs. As shown in the Supplemental Material of \cite{key-10},
we should note that the isotropic Raman response predominates over
the anisotropic response due to the molecular symmetry of the $\textrm{SiO}_{2}$.

However, our scenario omits the polarization effects. Therefore, taking
into account this point and using Eqs.\,(\ref{eq:2.30}), (\ref{eq:2.34})
and (\ref{eq:2.35}), it is straightforward to obtain the complex
amplitude of the instantaneous nonlinear polarization in the time
domain: 
\begin{align}
P_{\omega_{0}}^{\left(3\textrm{I}\right)} & =\varepsilon_{0}\gamma_{\textrm{I}}\left|E_{\omega_{0}}\right|^{2}E_{x,\omega_{0}},\label{eq:2.40}
\end{align}
where $\gamma_{\textrm{I}}:=\left(3/4\right)\chi_{\textrm{NL}}\left(1-f_{\textrm{R}}\right)=1.5\cdot10^{-22}\textrm{\ m}^{2}/\textrm{V}^{2}$
\textcolor{black}{at the wavelength of 1550 nm. The te}rm $3/4$ is
inferred from Eq.\,(\ref{eq:2.34}) by using the intrinsic permutation
symmetry of $\chi_{ijkl}^{\left(3\right)}$. A more detailed description
of this point can be found in \cite{key-8}.

Furthermore, the complex amplitude of the nonlinear polarization modeling
the nuclei motions (Raman response) can also be found by using Eqs.\,(\ref{eq:2.30}),
(\ref{eq:2.34}) and (\ref{eq:2.36}):
\begin{align}
P_{\omega_{0}}^{\left(3\textrm{R}\right)}\left(\boldsymbol{\boldsymbol{r}},t\right) & =\frac{1}{4}\varepsilon_{0}\chi_{\textrm{NL}}f_{\textrm{R}}\left\{ 2E_{\omega_{0}}\left(\boldsymbol{\boldsymbol{r}},t\right)\int_{-\infty}^{\infty}\left(h+u\right)\left(t-\zeta\right)\left|E_{\omega_{0}}\left(\boldsymbol{\boldsymbol{r}},\zeta\right)\right|^{2}\textrm{d}\zeta\right.\nonumber \\
 & +\left.E_{-\omega_{0}}\left(\boldsymbol{\boldsymbol{r}},t\right)\exp\left(-j2\omega_{0}t\right)\int_{-\infty}^{\infty}\left(h+u\right)\left(t-\zeta\right)E_{\omega_{0}}^{2}\left(\boldsymbol{\boldsymbol{r}},\zeta\right)\exp\left(j2\omega_{0}\zeta\right)\textrm{d}\zeta\right\} .\label{eq:2.41}
\end{align}
including in this case the independent variables to clarify the mathematical
discussion. In order to simplify Eq.\,(\ref{eq:2.41}), we can use
the commutation property of the convolution. In this way, the last
term of the RHS can be expressed as ($f:=h+u$):

\begin{align}
 & E_{-\omega_{0}}\left(\boldsymbol{\boldsymbol{r}},t\right)\exp\left(-j2\omega_{0}t\right)\int_{-\infty}^{\infty}f\left(t-\zeta\right)E_{\omega_{0}}^{2}\left(\boldsymbol{\boldsymbol{r}},\zeta\right)\exp\left(j2\omega_{0}\zeta\right)\textrm{d}\zeta\nonumber \\
 & \ \ =E_{-\omega_{0}}\left(\boldsymbol{\boldsymbol{r}},t\right)\exp\left(-j2\omega_{0}t\right)\int_{-\infty}^{\infty}f\left(\zeta\right)E_{\omega_{0}}^{2}\left(\boldsymbol{\boldsymbol{r}},t-\zeta\right)\exp\left(j2\omega_{0}\left(t-\zeta\right)\right)\textrm{d}\zeta\nonumber \\
 & \ \ =E_{-\omega_{0}}\left(\boldsymbol{\boldsymbol{r}},t\right)\int_{-\infty}^{\infty}f\left(\zeta\right)\exp\left(-j2\omega_{0}\zeta\right)E_{\omega_{0}}^{2}\left(\boldsymbol{\boldsymbol{r}},t-\zeta\right)\textrm{d}\zeta\nonumber \\
 & \ \ =E_{-\omega_{0}}\left(\boldsymbol{\boldsymbol{r}},t\right)\left[\left(f\left(t\right)\exp\left(-j2\omega_{0}t\right)\right)\ast E_{\omega_{0}}^{2}\left(\boldsymbol{\boldsymbol{r}},t\right)\right].\label{eq:2.42}
\end{align}
As can be noted, Eq.\,(\ref{eq:2.42}) involves a convolution of
the modulated Raman response with the complex amplitudes of the electric
field strength. Considering that the bandwidth of the Raman response
(given by the $f$ function) is around 15 THz \cite{key-9}, centered
at $-2\omega_{0}\approx-380$ THz, and the bandwidth of the complex
amplitudes is lower than 100 THz for ultra-short pulses wider than
10 fs (in baseband $\Omega:=\omega-\omega_{0}$), the Fourier transform
of the convolution is found to be null:
\begin{align}
\mathcal{F}\left[\left(f\left(t\right)\exp\left(-j2\omega_{0}t\right)\right)\ast E_{\omega_{0}}^{2}\left(\boldsymbol{\boldsymbol{r}},t\right)\right]=\ \ \ \ \ \ \ \ \ \ \ \ \ \ \ \ \ \nonumber \\
=\frac{1}{2\pi}\widetilde{F}\left(\Omega+2\omega_{0}\right)\left[\widetilde{E}_{\omega_{0}}\left(\boldsymbol{\boldsymbol{r}},\Omega\right)\ast\widetilde{E}_{\omega_{0}}\left(\boldsymbol{\boldsymbol{r}},\Omega\right)\right] & =0,\label{eq:2.43}
\end{align}
and therefore, the last term of the RHS of Eq.\,(\ref{eq:2.41})
can be neglected. Finally, defining the nonlinear constant $\gamma_{\textrm{R}}:=0.5\chi_{\textrm{NL}}f_{\textrm{R}}=3.2\cdot10^{-23}\textrm{\ m}^{2}/\textrm{V}^{2}$
\textcolor{black}{at the wavelength of 1550 nm, }the complex amplitude
of the nonlinear polarization modeling both electronic and nuclei
responses is found to be:
\begin{align}
P_{\omega_{0}}^{\left(3\right)}\left(\boldsymbol{\boldsymbol{r}},t\right) & =\varepsilon_{0}\gamma_{\textrm{I}}\left|E_{\omega_{0}}\left(\boldsymbol{\boldsymbol{r}},t\right)\right|^{2}E_{\omega_{0}}\left(\boldsymbol{\boldsymbol{r}},t\right)+\varepsilon_{0}\gamma_{\textrm{R}}\left(f\left(t\right)\ast\left|E_{\omega_{0}}\left(\boldsymbol{\boldsymbol{r}},t\right)\right|^{2}\right)E_{\omega_{0}}\left(\boldsymbol{\boldsymbol{r}},t\right).\label{eq:2.44}
\end{align}

Once we have derived the complex amplitude of the nonlinear polarization
in the time domain, Eq.\,(\ref{eq:2.29}) can be completed using
Eqs.\,(\ref{eq:2.32}) and (\ref{eq:2.44}). At this point, note
that the nonlinear constants $\gamma_{\textrm{I}}$ and $\gamma_{\textrm{R}}$
are also frequency dependent when operating with ultra-short optical
pulses in the femtosecond regime. However, we assume that the frequency
variation of these parameters is much lower than their average value
in the pulse bandwidth, as same as the frequency changes of $F$.
Hence, Eq.\,(\ref{eq:2.29}) becomes (for the sake of simplicity,
the independent variables are only included in the convolution operations
of the nonlinear terms): 
\begin{align}
F\exp\left(-j\beta_{0}z\right)\left[\partial_{z}^{2}\widetilde{A}-\left(j2\beta_{0}+\alpha\right)\left(\partial_{z}\widetilde{A}+\mathfrak{D}\widetilde{A}\right)\right]\nonumber \\
+\frac{\omega^{2}}{c_{0}^{2}}\gamma_{\textrm{I}}\exp\left(-\alpha z\right)\exp\left(-j\beta_{0}z\right)F^{3}\mathcal{F}\left\{ A\left|A\right|^{2}\exp\left(j\omega_{0}t\right)\right\} \nonumber \\
+\frac{\omega^{2}}{c_{0}^{2}}\gamma_{\textrm{R}}\exp\left(-\alpha z\right)\exp\left(-j\beta_{0}z\right)F^{3}\mathcal{F}\left\{ A\left[f\left(t\right)\ast\left|A\left(z,t\right)\right|^{2}\right]\exp\left(j\omega_{0}t\right)\right\}  & =0.\label{eq:2.45}
\end{align}

From Eq.\,(\ref{eq:2.45}), the PPE can be found by multiplying by
$F\exp(j\beta_{0}z)$ and integrating in an infinite cross-sectional
area of the fiber: 
\begin{align}
j\left(\frac{j}{2\beta_{0}}\partial_{z}^{2}+\partial_{z}+\mathfrak{D}\right)\widetilde{A} & =\exp\left(-\alpha z\right)\left\{ \widetilde{q}^{\left(\textrm{I}\right)}\mathcal{F}\left\{ A\left|A\right|^{2}\exp\left(j\omega_{0}t\right)\right\} \right.\nonumber \\
 & \left.+\widetilde{q}^{\left(\textrm{R}\right)}\mathcal{F}\left\{ A\left[f\left(t\right)\ast\left|A\left(z,t\right)\right|^{2}\right]\exp\left(j\omega_{0}t\right)\right\} \right\} ,\label{eq:2.46}
\end{align}
where $\widetilde{q}^{\left(\textrm{I}\right)}$ and $\widetilde{q}^{\left(\textrm{R}\right)}$
are the nonlinear mode-coupling coefficients (MCCs) defined as:
\begin{align}
\widetilde{q}^{\left(\textrm{S}\right)}\left(\omega\right) & :=\frac{\omega^{2}}{2c_{0}^{2}\beta_{0}}\frac{\iint_{\infty}\gamma_{\textrm{S}}\left(\omega\right)F^{4}\left(x,y,\omega\right)\textrm{d}x\textrm{d}y}{\iint_{\infty}F^{2}\left(x,y,\omega\right)\textrm{d}x\textrm{d}y};\ \ \textrm{S}\in\left\{ \textrm{I},\textrm{R}\right\} .\label{eq:2.47}
\end{align}
\textcolor{black}{Not}e that the frequency dependence of the nonlinear
MCCs also involves the nonlinear dispersion of $\gamma_{\textrm{I}}$
and $\gamma_{\textrm{R}}$, induced by the frequency dependence of
$n_{\textrm{NL}}$. In such a scenario, the function $n_{\textrm{NL}}$
can be approximated by a first-order Taylor series expansion as $n_{\textrm{NL}}\left(\omega\right)\simeq n_{\textrm{NL},0}+\left(\omega-\omega_{0}\right)n_{\textrm{NL},1}$,
where $n_{\textrm{NL},0}=2.6\cdot10^{-20}\textrm{\ m}^{2}/\textrm{W}$
and $n_{\textrm{NL},1}=8.3\cdot10^{-24}\textrm{\ ps·m}^{2}/\textrm{W}$
in silica fibers \cite{key-10}. Now, taking into account that $(1/2\beta_{0})\left|\partial_{z}^{2}\widetilde{A}\right|\ll\left|\partial_{z}\widetilde{A}\right|$
as demonstrated in \cite{key-7}, redefining the complex envelopes
in the frequency domain as $\widetilde{A}:=\widetilde{\mathcal{A}}\exp\left(\alpha z/2\right)$
and assuming the attenuation coefficient $\alpha$ with low frequency
dependence satisfying $\alpha\left(\omega_{0}\right)\gg\textrm{d}\alpha\left(\omega=\omega_{0}\right)/\textrm{d}\omega$,
Eq.\,(\ref{eq:2.46}) can be reduced to:
\begin{align}
j\left(\partial_{z}+\mathfrak{D}+\frac{\alpha}{2}\right)\widetilde{\mathcal{A}} & =\widetilde{q}^{\left(\textrm{I}\right)}\mathcal{F}\left\{ \mathcal{A}\left|\mathcal{A}\right|^{2}\exp\left(j\omega_{0}t\right)\right\} +\widetilde{q}^{\left(\textrm{R}\right)}\mathcal{F}\left\{ \mathcal{A}\left[f\left(t\right)\ast\left|\mathcal{A}\left(z,t\right)\right|^{2}\right]\exp\left(j\omega_{0}t\right)\right\} .\label{eq:2.48}
\end{align}

In order to derive the final expression of the PPE in the time domain,
we should perform the following Taylor series expansion at $\omega=\omega_{0}$
of the holomorphic functions (we denote the instantaneous and Raman
nonlinear MCCs by the superindex $\textrm{S}\in\left\{ \textrm{I},\textrm{R}\right\} $):
\begin{align}
\widetilde{q}^{\left(\textrm{S}\right)}\left(\omega\right) & =\sum_{n=0}^{\infty}\frac{1}{n!}\left(\omega-\omega_{0}\right)^{n}\left.\frac{\textrm{d}^{n}\widetilde{q}^{\left(\textrm{S}\right)}\left(\omega\right)}{\textrm{d}\omega^{n}}\right\rfloor _{\omega_{0}}\equiv\sum_{n=0}^{\infty}\frac{1}{n!}\left(\omega-\omega_{0}\right)^{n}\widetilde{q}_{n}^{\left(\textrm{S}\right)};\label{eq:2.49}\\
\alpha\left(\omega\right) & \simeq\alpha\left(\omega_{0}\right)+\left(\omega-\omega_{0}\right)\left.\frac{\textrm{d}\alpha\left(\omega\right)}{\textrm{d}\omega}\right\rfloor _{\omega_{0}}\equiv\alpha_{0}+\left(\omega-\omega_{0}\right)\alpha_{1},\label{eq:2.50}
\end{align}
which can be expressed in the time domain by the following linear
operators:
\begin{equation}
\hat{\textrm{q}}^{\left(\textrm{S}\right)}:=\sum_{n=0}^{\infty}\frac{\left(-j\right)^{n}}{n!}\widetilde{q}_{n}^{\left(\textrm{S}\right)}\partial_{t}^{n}\ \ \ \ \ \ \ \widehat{\mathrm{\alpha}}:=\alpha_{0}-j\alpha_{1}\partial_{t},\label{eq:2.51}
\end{equation}
and the complex function $\mathfrak{D}$ is expressed in the time
domain by \textcolor{black}{the dispersion operator: }
\begin{equation}
\widehat{\textrm{D}}:=\sum_{n=1}^{\infty}\frac{\left(-j\right)^{n-1}}{n!}\beta_{n}\partial_{t}^{n}.\label{eq:2.52}
\end{equation}
Finally, applying the inverse Fourier transform to Eq.\,(\ref{eq:2.48})
the final expression of the PPE is derived in the time domain:\\

\noindent\fbox{\begin{minipage}[t]{1\columnwidth - 2\fboxsep - 2\fboxrule}%
\begin{align}
j\left(\partial_{z}+\widehat{\textrm{D}}+\frac{1}{2}\widehat{\alpha}\right)\mathcal{A}\left(z,t\right) & =\hat{\textrm{q}}^{\left(\textrm{I}\right)}\left(\left|\mathcal{A}\left(z,t\right)\right|^{2}\mathcal{A}\left(z,t\right)\right)+\hat{\textrm{q}}^{\left(\textrm{R}\right)}\left[\left(f\left(t\right)\ast\left|\mathcal{A}\left(z,t\right)\right|^{2}\right)\mathcal{A}\left(z,t\right)\right].\label{eq:2.53}
\end{align}

\vspace{0.2cm}
\end{minipage}}

\pagebreak{}

\noindent From this equation, the following remarks are in order:
\begin{itemize}
\item In order to reduce the computational complexity of the PPE, $\widehat{\textrm{D}}$
and $\hat{\textrm{q}}^{\left(\textrm{I},\textrm{R}\right)}$ can be
calculated in optical fibers by using a third- and first-order Taylor
series expansion, respectively \cite{key-10}.
\item The units of $\mathcal{A}\left(z,t\right)$ are V/m. If we are interested
in the optical power propagated by the optical pulse in the $\textrm{LP}_{01}$
mode, we must integrate the time average of the Poynting vector ($\boldsymbol{\mathcal{S}}=\boldsymbol{\mathcal{E}}\times\boldsymbol{\mathcal{H}}$)
in a period of the optical carrier ($T_{0}=2\pi/\omega_{0}$) and
in an infinite cross-sectional area of the fiber, which can be expressed
as a function of the complex envelope as follows:\footnote{Note that the complex envelope is assumed to be constant in a period
of the optical carrier. Thus, in this case, the time average operator
only applies over the rapidly-varying exponential terms of $\boldsymbol{\mathcal{E}}$
and $\boldsymbol{\mathcal{H}}$.} 
\begin{equation}
P\left(z,t\right)=\iint_{\infty}\left\langle \boldsymbol{\mathcal{E}}\times\boldsymbol{\mathcal{H}}\right\rangle \cdot\hat{z}\textrm{d}x\textrm{d}y=\mathscr{C}^{\left(\textrm{P}\right)}\left|\mathcal{A}\left(z,t\right)\right|^{2},\label{eq:2.54}
\end{equation}
where:\footnote{In the single-mode regime $F$ is a real function, that is, $F=F^{*}$.}
\begin{equation}
\mathscr{C}^{\left(\textrm{P}\right)}:=\iint_{\infty}\frac{1}{2\eta\left(x,y,\omega_{0}\right)}F^{2}\left(x,y,\omega_{0}\right)\textrm{d}x\textrm{d}y,\label{eq:2.55}
\end{equation}
with $\eta$ the intrinsic impedance of the medium, which can be calculated
as a function of the intrinsic impedance in vacuum $\eta_{0}=120\pi$
($\Omega$) as $\eta\left(x,y,\omega_{0}\right)=\eta_{0}/n_{\textrm{co}}\left(x,y,\omega_{0}\right)$
in the core and $\eta\left(x,y,\omega_{0}\right)=\eta_{0}/n_{\textrm{cl}}\left(x,y,\omega_{0}\right)$
in the cladding.
\end{itemize}

\section{Pulse propagation in uniform dielectric media\label{sec:3}}

Let us now consider other basic scenario of nonlinear optics: a pulsed
plane wave propagating along the $z$ axis through an uniform (or
unguided\footnote{The terminology \emph{unguided} medium can also be found in the literature
as \emph{uniform} medium (homogeneous refractive index). This concept
is intimately related in optical communications with a \emph{wireless}
propagation.}) nonlinear dielectric medium. The mathematical derivation of the
PPE is similar to the previous section. However, the following points
should be taken into account in each step depicted in Fig.\,2.

\subsubsection*{First step: Ansatz of the electric field strength of the medium}

The ansatz of Maxwell's equations of a pulsed plane wave propagating
in an uniform medium is similar to an optical pulse propagating through
an optical fiber. Nevertheless, the weakly-guiding approximation is
not required in this case. Without loss of generality, we can consider
only a single component of the electric field strength when assuming
a pulsed plane wave propagated through an uniform ($\equiv$ homogeneous)
isotropic optical medium. In this way, the ansatz of step 1 is described
by the following equations in the time domain: 
\begin{align}
\boldsymbol{\mathcal{E}}\left(\boldsymbol{\boldsymbol{r}},t\right) & =\mathcal{E}\left(\boldsymbol{\boldsymbol{r}},t\right)\hat{\tau}\simeq\textrm{Re}\left\{ E_{\omega_{0}}\left(\boldsymbol{\boldsymbol{r}},t\right)\exp\left(j\omega_{0}t\right)\right\} \hat{\tau};\label{eq:3.1}\\
E_{\omega_{0}}\left(\boldsymbol{\boldsymbol{r}},t\right) & =\frac{1}{2\pi}\int_{-\infty}^{\infty}\widetilde{A}\left(\boldsymbol{\boldsymbol{r}},\omega-\omega_{0}\right)\exp\left(-j\beta_{0}z-\frac{1}{2}\alpha\left(\omega\right)z\right)\exp\left(j\left(\omega-\omega_{0}\right)t\right)\textrm{d}\omega,\label{eq:3.2}
\end{align}
and in the frequency domain: 
\begin{align}
\widetilde{\boldsymbol{\mathcal{E}}}\left(\boldsymbol{\boldsymbol{r}},\omega\right) & \simeq\textrm{Re}\left\{ \widetilde{E}_{\omega_{0}}\left(\boldsymbol{\boldsymbol{r}},\omega-\omega_{0}\right)\right\} \hat{\tau};\label{eq:3.3}\\
\widetilde{E}_{\omega_{0}}\left(\boldsymbol{\boldsymbol{r}},\omega-\omega_{0}\right) & =\widetilde{A}\left(\boldsymbol{\boldsymbol{r}},\omega-\omega_{0}\right)\exp\left(-j\beta_{0}z-\frac{1}{2}\alpha\left(\omega\right)z\right).\label{eq:3.4}
\end{align}
It should be noted that we have included the spatial dependence of
$\widetilde{A}$ not only in the $z$ axis, but also in the transverse
coordinates $\left(x,y\right)$ to include the diffraction of the
optical pulse during the propagation. In addition, note that $F\equiv1$
and $\beta_{0}\equiv k_{z}\left(\omega_{0}\right)$ in unguided media,
where $k_{z}$ is the $z$-component of the wavevector. 

\subsubsection*{Second step: Wave equation of the guided mode}

There is not a guided mode ($\equiv$ bound state). Thus, the second
step {[}Eq.\,(\ref{eq:2.13}){]} is irrelevant in this case. We do
not need to replace a term of the form $\triangle_{\textrm{T}}F$
{[}Eq.\,(\ref{eq:2.28}){]} in the fourth step.

\subsubsection*{Third step: Nonlinear complex amplitude wave equation}

The nonlinear complex amplitude wave equation is found to be the same
as Eq.\,(\ref{eq:2.23}) given that the linear constitutive relation
$\boldsymbol{\mathcal{P}}^{\left(1\right)}$-$\boldsymbol{\mathcal{E}}$
is the same as in an optical fiber. Nonetheless, in this case it is
more convenient to retain the imaginary part of $\chi^{\left(1\right)}$.
In silica fibers, $\textrm{Im}\left\{ \chi^{\left(1\right)}\right\} $
can be neglected in the third transmission window because we operate
far from the resonance frequencies of the medium. Unfortunately, this
assumption is not general and should be discarded in this scenario.
In addition, accordingly with the unguided nature of the optical propagation,
we should also consider a homogeneous $\chi^{\left(1\right)}$ distribution.
As a result, the nonlinear complex amplitude wave equation is found
to be: 
\begin{equation}
\left(\triangle+\frac{\omega^{2}}{c_{0}^{2}}\widetilde{\varepsilon}_{\textrm{r}}\left(\omega\right)\right)\widetilde{E}_{\omega_{0}}\left(\boldsymbol{\boldsymbol{r}},\omega-\omega_{0}\right)=-\omega^{2}\mu_{0}\widetilde{P}_{\omega_{0}}^{\left(3\right)}\left(\boldsymbol{\boldsymbol{r}},\omega-\omega_{0}\right),\label{eq:3.5}
\end{equation}
with: 
\begin{equation}
\frac{\omega^{2}}{c_{0}^{2}}\widetilde{\varepsilon}_{\textrm{r}}\left(\omega\right)=\frac{\omega^{2}}{c_{0}^{2}}\left(1+\widetilde{\chi}^{\left(1\right)}\left(\omega\right)\right)\equiv\left(\beta\left(\omega\right)-j\frac{1}{2}\alpha\left(\omega\right)\right)^{2},\label{eq:3.6}
\end{equation}
and $\beta\left(\omega\right)=\beta_{0}-j\mathfrak{D}\left(\omega\right)$.

\subsubsection*{Fourth step: Complex envelope propagation equation}

Following the same procedure as in the fourth step in optical fibers,
i.e., replacing the results of step 1 {[}Eq.\,(\ref{eq:3.4}){]}
in step 3 {[}Eq.\,(\ref{eq:3.5}){]}, we will be able to derive the
PPE in unguided media. More specifically, Eq.\,(\ref{eq:2.29}) becomes:
\begin{equation}
\exp\left(-j\beta_{0}z\right)\left[\triangle_{\textrm{T}}+\partial_{z}^{2}-\left(j2\beta_{0}+\alpha\right)\left(\partial_{z}+\mathfrak{D}\right)\right]\widetilde{A}=-\omega^{2}\mu_{0}\widetilde{P}_{\omega_{0}}^{\left(3\right)}\exp\left(\frac{1}{2}\alpha z\right),\label{eq:3.7}
\end{equation}
with a new term $\triangle_{\textrm{T}}\widetilde{A}$ accounting
for the diffraction of the complex envelope. Once again, if we assume
that the pulse is propagated through a dielectric media with electronic
absorption resonances well above the frequencies of the optical fields,
$\widetilde{P}_{\omega_{0}}^{\left(3\right)}$ can be written as in
optical fibers {[}Eq.\,(\ref{eq:2.34}){]}: an instantaneous response
accounting for the electronic and a delayed response induced by the
nuclei motions \cite{key-3}. In addition, note that the term $\widetilde{P}_{3\omega_{0}}^{\left(3\right)}$
can also be neglected if we consider that $\left|\textrm{Re}\left\{ \widetilde{\chi}^{\left(1\right)}\left(3\omega\right)-\widetilde{\chi}^{\left(1\right)}\left(\omega\right)\right\} \right|\gg0$
due to the temporal dispersive nature of the optical medium \cite{key-8}.
All in all, the PPE is finally derived of the form:

\noindent\fbox{\begin{minipage}[t]{1\columnwidth - 2\fboxsep - 2\fboxrule}%
\begin{align}
j\left(\frac{j}{2\beta_{0}}\triangle_{\textrm{T}}+\partial_{z}+\widehat{\textrm{D}}+\frac{1}{2}\widehat{\alpha}\right)\mathcal{A}\left(\boldsymbol{\boldsymbol{r}},t\right) & =\hat{\textrm{q}}^{\left(\textrm{I}\right)}\left(\left|\mathcal{A}\left(\boldsymbol{\boldsymbol{r}},t\right)\right|^{2}\mathcal{A}\left(\boldsymbol{\boldsymbol{r}},t\right)\right)\nonumber \\
 & +\hat{\textrm{q}}^{\left(\textrm{R}\right)}\left[\left(f\left(t\right)\ast\left|\mathcal{A}\left(\boldsymbol{\boldsymbol{r}},t\right)\right|^{2}\right)\mathcal{A}\left(\boldsymbol{\boldsymbol{r}},t\right)\right].\label{eq:3.8}
\end{align}

\vspace{0.2cm}
\end{minipage}}\pagebreak{}

\noindent Along this line, the following points should be taken into
account:
\begin{itemize}
\item The linear operators $\hat{\textrm{q}}^{\left(\textrm{I}\right)}$
and $\hat{\textrm{q}}^{\left(\textrm{R}\right)}$ describe the frequency
dependence of the nonlinear MCCs $\widetilde{q}^{\left(\textrm{I}\right)}\left(\omega\right)$
and $\widetilde{q}^{\left(\textrm{R}\right)}\left(\omega\right)$
in the time domain, respectively. However, given that $F\equiv1$,
Eq.\,(\ref{eq:2.47}) becomes: 
\begin{equation}
\widetilde{q}^{\left(\textrm{S}\right)}\left(\omega\right)=\frac{\omega^{2}\gamma_{\textrm{S}}\left(\omega\right)}{2c_{0}^{2}\beta_{0}};\ \ \textrm{S}\in\left\{ \textrm{I},\textrm{R}\right\} .\label{eq:3.9}
\end{equation}
\item The term $(j/2\beta_{0})\partial_{z}^{2}\mathcal{A}$ has been neglected
in the LHS of the PPE as in the previous section because $(1/2\beta_{0})\left|\partial_{z}^{2}\mathcal{A}\right|\ll\left|\partial_{z}\mathcal{A}\right|$.
Nevertheless, the term $(j/2\beta_{0})\triangle_{\textrm{T}}\mathcal{A}$
cannot be neglected because, in general, $(1/2\beta_{0})\left|\triangle_{\textrm{T}}\mathcal{A}\right|$
can be found of the same order (or higher) as $\left|\partial_{z}\mathcal{A}\right|$.
Remarkably, this term accounts for the pulse diffraction, as commented
before.
\item The linear operators $\widehat{\textrm{D}}$ and $\hat{\textrm{q}}^{\left(\textrm{I},\textrm{R}\right)}$
can also be calculated by using a \emph{finite} Taylor series expansion.
Unfortunately, the required order of the Taylor series in each term
is not general. It depends on the particular dielectric medium simulated
in the numerical calculations (see \cite{key-11} for more details). 
\item The units of $\mathcal{A}\left(\boldsymbol{\boldsymbol{r}},t\right)$
are V/m. If we are interested in the optical \emph{intensity} (W/m\textsuperscript{2})
propagated by the pulsed plane wave, we must calculate the module
of the time average of the Poynting vector in a period of the optical
carrier ($T_{0}=2\pi/\omega_{0}$):\footnote{As in the previous section, the complex envelope is assumed to be
constant in a period of the optical carrier. Thus, the time average
operator only applies over the rapidly-varying exponential terms of
$\boldsymbol{\mathcal{E}}$ and $\boldsymbol{\mathcal{H}}$.} 
\begin{equation}
I\left(\boldsymbol{\boldsymbol{r}},t\right)=\left\Vert \left\langle \boldsymbol{\mathcal{S}}\right\rangle \right\Vert =\left\langle \boldsymbol{\mathcal{E}}\times\boldsymbol{\mathcal{H}}\right\rangle \cdot\hat{z}=\frac{1}{2\eta\left(\omega_{0}\right)}\left|\mathcal{A}\left(\boldsymbol{\boldsymbol{r}},t\right)\right|^{2},\label{eq:3.10}
\end{equation}
with $\eta\left(\omega_{0}\right)=\eta_{0}/n\left(\omega_{0}\right)$
and $n\left(\omega_{0}\right):=\sqrt{1+\textrm{Re}\left\{ \widetilde{\chi}^{\left(1\right)}\left(\omega_{0}\right)\right\} }$.
Of course, we do not calculate the optical power of a pulsed plane
wave because is infinite (note that a plane wave impinges an infinite
transverse plane).
\end{itemize}

\section{White Light Continuum (WLC) in gases\label{sec:4}}

Following \cite{key-12} as a guideline, it should be noted that the
ideal phase constant $\beta\equiv k_{z}$ of Section\,\ref{sec:3}
is \emph{perturbed} by the presence of the plasma. Therefore, we can
use the same approach as in \cite{key-7,key-10}, i.e., modeling the
plasma as a \emph{perturbation} of the optical medium. To clarify
this approach, let us perform once again a detailed review of the
derivation of the PPE from Maxwell's equations. The steps are similar
to Fig.\,2.

\subsubsection*{First step: Ansatz of the electric field strength of the medium}

The ansatz of Maxwell's equations of a pulsed plane wave propagating
in this scenario is similar to the ansatz of the previous section.
Nevertheless, we should include the perturbation of the plasma in
the propagation constant $\gamma=\alpha/2+j\beta$. To this end, we
should assume the electric field strength in the time domain of the
form:
\begin{align}
\boldsymbol{\mathcal{E}}\left(\boldsymbol{\boldsymbol{r}},t\right) & =\mathcal{E}\left(\boldsymbol{\boldsymbol{r}},t\right)\hat{\tau}\simeq\textrm{Re}\left\{ E_{\omega_{0}}\left(\boldsymbol{\boldsymbol{r}},t\right)\exp\left(j\omega_{0}t\right)\right\} \hat{\tau};\label{eq:4.1}\\
E_{\omega_{0}}\left(\boldsymbol{\boldsymbol{r}},t\right) & =\frac{1}{2\pi}\int_{-\infty}^{\infty}\widetilde{A}\left(\boldsymbol{\boldsymbol{r}},\omega-\omega_{0}\right)\exp\left(-j\beta_{0}^{\left(\textrm{eq}\right)}z-\frac{1}{2}\alpha\left(\omega\right)z\right)\exp\left(j\left(\omega-\omega_{0}\right)t\right)\textrm{d}\omega,\label{eq:4.2}
\end{align}
and in the frequency domain: 
\begin{align}
\widetilde{\boldsymbol{\mathcal{E}}}\left(\boldsymbol{\boldsymbol{r}},\omega\right) & \simeq\textrm{Re}\left\{ \widetilde{E}_{\omega_{0}}\left(\boldsymbol{\boldsymbol{r}},\omega-\omega_{0}\right)\right\} \hat{\tau};\label{eq:4.3}\\
\widetilde{E}_{\omega_{0}}\left(\boldsymbol{\boldsymbol{r}},\omega-\omega_{0}\right) & =\widetilde{A}\left(\boldsymbol{\boldsymbol{r}},\omega-\omega_{0}\right)\exp\left(-j\beta_{0}^{\left(\textrm{eq}\right)}z-\frac{1}{2}\alpha\left(\omega\right)z\right).\label{eq:4.4}
\end{align}
with $\beta_{0}^{\left(\textrm{eq}\right)}:=\beta^{\left(\textrm{eq}\right)}\left(\omega=\omega_{0}\right)$,
where $\beta^{\left(\textrm{eq}\right)}\left(\omega\right)$ is the
\emph{equivalent} phase constant at each spectral component $\omega$
accounting for the ideal phase constant $\beta\left(\omega\right)$
and the perturbation induced by the plasma $\beta^{\left(\textrm{PL}\right)}\left(\omega\right)$:
\begin{equation}
\beta^{\left(\textrm{eq}\right)}\left(\omega\right):=\beta\left(\omega\right)+\beta^{\left(\textrm{PL}\right)}\left(\omega\right).\label{eq:4.5}
\end{equation}

\subsubsection*{Second step: Wave equation of the guided mode}

There is not a guided mode ($\equiv$ bound state). Therefore, the
second step {[}Eq.\,(\ref{eq:2.13}){]} is irrelevant in this case.
We do not need to replace a term of the form $\triangle_{\textrm{T}}F$
{[}Eq.\,(\ref{eq:2.28}){]} in the fourth step. Here, we have the
same conclusions as in the second step of Section\,\ref{sec:3}.

\subsubsection*{Third step: Nonlinear complex amplitude wave equation}

The nonlinear complex amplitude wave equation is found to be the same
as Eq.\,(\ref{eq:3.5}) given that the linear constitutive relation
$\boldsymbol{\mathcal{P}}^{\left(1\right)}$-$\boldsymbol{\mathcal{E}}$
is the same as in the previous section. We repeat this equation for
clarity: 
\begin{equation}
\left(\triangle+\frac{\omega^{2}}{c_{0}^{2}}\widetilde{\varepsilon}_{\textrm{r}}\left(\omega\right)\right)\widetilde{E}_{\omega_{0}}\left(\boldsymbol{\boldsymbol{r}},\omega-\omega_{0}\right)=-\omega^{2}\mu_{0}\widetilde{P}_{\omega_{0}}^{\left(3\right)}\left(\boldsymbol{\boldsymbol{r}},\omega-\omega_{0}\right),\label{eq:4.6}
\end{equation}
with: 
\begin{equation}
\frac{\omega^{2}}{c_{0}^{2}}\widetilde{\varepsilon}_{\textrm{r}}\left(\omega\right)=\frac{\omega^{2}}{c_{0}^{2}}\left(1+\widetilde{\chi}^{\left(1\right)}\left(\omega\right)\right)\equiv\left(\beta^{\left(\textrm{eq}\right)}\left(\omega\right)-j\frac{1}{2}\alpha\left(\omega\right)\right)^{2}.\label{eq:4.7}
\end{equation}
It is worthy to note that $\widetilde{\varepsilon}_{\textrm{r}}$
involves now the contribution of the plasma via $\beta^{\left(\textrm{eq}\right)}$.

\subsubsection*{Fourth step: Complex envelope propagation equation}

Following the same procedure as in the fourth step of previous examples,
we should replace the results of step 1 {[}Eq.\,(\ref{eq:4.4}){]}
in step 3 {[}Eq.\,(\ref{eq:4.6}){]}. Thus, in this case Eq.\,(\ref{eq:3.7})
becomes: 
\begin{equation}
\exp\left(-j\beta_{0}z\right)\left[\triangle_{\textrm{T}}+\partial_{z}^{2}-\left(j2\beta_{0}+\alpha\right)\left(\partial_{z}+j\beta_{0}^{\left(\textrm{PL}\right)}+\mathfrak{D}^{\left(\textrm{eq}\right)}\right)\right]\widetilde{A}=-\omega^{2}\mu_{0}\widetilde{P}_{\omega_{0}}^{\left(3\right)}\exp\left(\frac{1}{2}\alpha z\right),\label{eq:4.8}
\end{equation}
with: 
\begin{equation}
\mathfrak{D}^{\left(\textrm{eq}\right)}\left(\omega\right):=\sum_{k=1}^{\infty}\frac{j}{k!}\left(\omega-\omega_{0}\right)^{k}\beta_{k}^{\left(\textrm{eq}\right)},\label{eq:4.9}
\end{equation}
and $\beta_{k}^{\left(\textrm{eq}\right)}:=\textrm{d}^{k}\beta^{\left(\textrm{eq}\right)}\left(\omega=\omega_{0}\right)/\textrm{d}\omega^{k}$.
According to \cite{key-12}, the term $\beta_{0}^{\left(\textrm{PL}\right)}:=\beta^{\left(\textrm{PL}\right)}\left(\omega_{0}\right)$
will be the predominant perturbation of the plasma during the propagation
of the optical pulse. The other terms $\beta_{k>0}^{\left(\textrm{PL}\right)}$,
included in $\mathfrak{D}^{\left(\textrm{eq}\right)}$, describe the
higher-order dispersion terms of the plasma. Concretely, the holomorphic
complex function $\mathfrak{D}^{\left(\textrm{eq}\right)}$ can be
modeled in the time domain by the equivalent dispersion operator $\widehat{\textrm{D}}^{(\textrm{eq})}$
defined as: 
\begin{equation}
\widehat{\textrm{D}}^{(\textrm{eq})}:=\sum_{k=1}^{\infty}\frac{\left(-j\right)^{k-1}}{k!}\beta_{k}^{(\textrm{eq})}\partial_{t}^{k}.\label{eq:4.10}
\end{equation}
Taking into account that $\beta_{k}^{(\textrm{eq})}=\beta_{k}+\beta_{k}^{(\textrm{PL})}$,
we can rewrite $\widehat{\textrm{D}}^{(\textrm{eq})}$ as the linear
contribution of the ideal dispersion operator $\widehat{\textrm{D}}$
and the higher-order dispersion terms of the plasma, modeled by $\widehat{\textrm{D}}^{(\textrm{PL})}$,
defined as: 
\begin{equation}
\widehat{\textrm{D}}^{(\textrm{PL})}:=\sum_{k=1}^{\infty}\frac{\left(-j\right)^{k-1}}{k!}\beta_{k}^{(\textrm{PL})}\partial_{t}^{k}.\label{eq:4.11}
\end{equation}
That is, we separate the dispersion induced by the frequency dependence
of the ideal phase constant and the plasma phase constant: $\widehat{\textrm{D}}^{(\textrm{eq})}=\widehat{\textrm{D}}+\widehat{\textrm{D}}^{(\textrm{PL})}$.

Now, if we assume that $\widetilde{P}_{\omega_{0}}^{\left(3\right)}$
can be written as in previous sections, i.e., with an instantaneous
response accounting for the electronic and a delayed response induced
by the nuclei motions, the PPE is derived in the time domain of the
form:\\

\noindent\fbox{\begin{minipage}[t]{1\columnwidth - 2\fboxsep - 2\fboxrule}%
\begin{align}
j\left(\frac{j}{2\beta_{0}}\triangle_{\textrm{T}}+\partial_{z}+\widehat{\textrm{D}}+\frac{1}{2}\widehat{\alpha}\right)\mathcal{A}\left(\boldsymbol{\boldsymbol{r}},t\right) & =\beta_{0}^{\left(\textrm{PL}\right)}\mathcal{A}\left(\boldsymbol{\boldsymbol{r}},t\right)-j\widehat{\textrm{D}}^{(\textrm{PL})}\mathcal{A}\left(\boldsymbol{\boldsymbol{r}},t\right)\nonumber \\
 & +\hat{\textrm{q}}^{\left(\textrm{I}\right)}\left(\left|\mathcal{A}\left(\boldsymbol{\boldsymbol{r}},t\right)\right|^{2}\mathcal{A}\left(\boldsymbol{\boldsymbol{r}},t\right)\right)\nonumber \\
 & +\hat{\textrm{q}}^{\left(\textrm{R}\right)}\left[\left(f\left(t\right)\ast\left|\mathcal{A}\left(\boldsymbol{\boldsymbol{r}},t\right)\right|^{2}\right)\mathcal{A}\left(\boldsymbol{\boldsymbol{r}},t\right)\right].\label{eq:4.12}
\end{align}

\vspace{0.2cm}
\end{minipage}}

\vspace{0.3cm}

\noindent Finally, from the above equation the following points are
in order:
\begin{itemize}
\item The linear operators $\hat{\textrm{q}}^{\left(\textrm{I}\right)}$
and $\hat{\textrm{q}}^{\left(\textrm{R}\right)}$ describe the frequency
dependence of the nonlinear MCCs $\widetilde{q}^{\left(\textrm{I}\right)}\left(\omega\right)$
and $\widetilde{q}^{\left(\textrm{R}\right)}\left(\omega\right)$
in the time domain, respectively. These nonlinear MCCs should be calculated
as indicated in Eq.\,(\ref{eq:3.9}).
\item The linear operators can be calculated by using a \emph{finite} Taylor
series expansion. The minimum order required to describe accurately
the optical medium and the plasma should be investigated.
\item The units of $\mathcal{A}\left(\boldsymbol{\boldsymbol{r}},t\right)$
are V/m. If we are interested in the optical \emph{intensity} (W/m\textsuperscript{2})
propagated by the pulsed plane wave, we can use Eq.\,(\ref{eq:3.10})
to this end. However, note that the intrinsic impedance $\eta\left(\omega_{0}\right)$
must be calculated taking into account the plasma contribution. Specifically,
the optical intensity is found to be: 
\begin{align}
I\left(\boldsymbol{\boldsymbol{r}},t\right) & =\frac{1}{2\eta\left(\omega_{0}\right)}\left|\mathcal{A}\left(\boldsymbol{\boldsymbol{r}},t\right)\right|^{2};\label{eq:4.13}\\
\eta\left(\omega_{0}\right) & =\frac{\eta_{0}}{\sqrt{1+\textrm{Re}\left\{ \widetilde{\chi}^{\left(1\right)}\left(\omega_{0}\right)\right\} }}=\frac{\eta_{0}}{n\left(\omega_{0}\right)+\Delta n^{\left(\textrm{PL}\right)}\left(\omega_{0}\right)}.\label{eq:4.14}
\end{align}
\end{itemize}
\pagebreak{}

\end{document}